\documentclass[twocolumn]{aastex63}

\usepackage{graphicx}	
\usepackage{amsmath}	
\usepackage{amssymb}	
\usepackage{bm}		

\submitjournal{ApJ}

\shorttitle{Anomaly Detection}
\shortauthors{Villar et al.}

\newcommand{\Ntrans}{1,129,184}
\newcommand{\pertde}{0.6}
\newcommand{\perii}{25.8}
\newcommand{\periin}{1.8}
\newcommand{\periax}{1.8}
\newcommand{\peribc}{5.8}
\newcommand{\perbg}{1.2}
\newcommand{\peragn}{6.3}
\newcommand{\peria}{54.2}
\newcommand{\perslsn}{2.2}
\newcommand{\perilot}{0.08}
\newcommand{\percart}{0.31}
\newcommand{\perpisn}{0.07}

\newcommand{\LikeL}{\mathcal{L}}
\newcommand{\plasticc}{PLAsTiCC}

\begin{document}

\title{A Deep Learning Approach for Active Anomaly Detection of Extragalactic Transients}

\correspondingauthor{V.~Ashley Villar}
\email{vav2110@columbia.edu}

\author[0000-0002-0786-7307]{V.~Ashley Villar}
\affiliation{Simons Junior Fellow, Department of Astronomy, Columbia University, New York, NY 10027, USA}
\author[0000-0002-6458-3423]{Miles Cranmer}
\affiliation{Princeton University,
   Princeton, NJ 08544, USA}
   
\author[0000-0002-9392-9681]{Edo~Berger}
\affiliation{Center for Astrophysics \textbar{} Harvard \& Smithsonian, 60 Garden Street, Cambridge, MA 02138-1516, USA}
\affiliation{The NSF AI Institute for Artificial Intelligence and Fundamental Interactions}

\author[0000-0002-3011-4784]{Gabriella Contardo}
\affiliation{Flatiron Institute,
   New York City, NY 10010, USA}

\author{Shirley Ho}
\affiliation{Flatiron Institute,
   New York City, NY 10010, USA}

\author[0000-0002-0832-2974]{Griffin~Hosseinzadeh}
\affiliation{Center for Astrophysics \textbar{} Harvard \& Smithsonian, 60 Garden Street, Cambridge, MA 02138-1516, USA}
\affiliation{The NSF AI Institute for Artificial Intelligence and Fundamental Interactions}

\author[0000-0003-0680-4838]{Joshua Yao-Yu Lin}
\affiliation{University of Illinois at Urbana-Champaign, Urbana, IL 61820, USA}

\begin{abstract}
There is a shortage of multi-wavelength and spectroscopic followup capabilities given the number of transient and variable astrophysical events discovered through wide-field, optical surveys such as the upcoming Vera C.~Rubin Observatory. From the haystack of potential science targets, astronomers must allocate scarce resources to study a selection of needles in real time. Here we present a variational recurrent autoencoder neural network to encode simulated Rubin Observatory extragalactic transient events using 1\% of the \plasticc\ dataset to train the autoencoder. Our unsupervised method uniquely works with unlabeled, real time, multivariate and aperiodic data. We rank \Ntrans\ events based on an anomaly score estimated using an isolation forest. We find that our pipeline successfully ranks rarer classes of transients as more anomalous. Using simple cuts in anomaly score and uncertainty, we identify a pure ($\approx95$\% pure) sample of rare transients (i.e., transients other than Type Ia, Type II and Type Ibc supernovae) including superluminous and pair-instability supernovae. Finally, our algorithm is able to identify these transients as anomalous well before peak, enabling real-time follow up studies in the era of the Rubin Observatory.

\end{abstract}


\section{Introduction}

Wide-field, optical surveys such as the Asteroid Terrestrial-impact Last Alert System (ATLAS; \citealt{jedicke2012atlas}), the All-Sky Automated Survey for SuperNovae (ASAS-SN; \citealt{shappee2014all}), the Panoramic Survey Telescope and Rapid Response System 1 (Pan-STARRS1; \citealt{2016arXiv161205560C}) and the Zwicky Transient Facility (ZTF; \citealt{bellm2018zwicky}) have exponentially increased the discovery rate of new transient events that vary on day to year timescales. The upcoming Vera C.~Rubin Observatory \citep{2019ApJ...873..111I} and its decade-long Legacy Survey of Space and Time (LSST) will greatly accelerate this discovery rate to millions of new transients annually. However, a limited fraction (likely $\lesssim0.1$\%) of all events can be followed up with dedicated spectroscopic and multi-wavelength campaigns. Identifying transients worthwhile of followup will be akin to finding needles in a haystack. Adding to the challenge, we will need to identify such events quickly to capture events pre- or near-peak to fully optimize the efficiency of follow-up campaigns.

Over the past few years there have been several initial efforts aimed at photometrically classifying transients to build pure samples of known transient classes \citep{boone2019avocado, muthukrishna2019rapid,pasquet2019pelican, hosseinzadeh2020photometric, sanchez2020alert,villar2020superraenn}. However, even the rarest transients known today, like superluminous supernovae (SLSNe) and tidal disruption events (TDEs), will be discovered by the thousands in the era of LSST (e.g., \citealt{villar2018superluminous,bricman2020prospects}). Detection and classification algorithms sensitive to anomalous transients are essential in order to discover unexpected and even rarer phenomena.

There is a growing literature on anomaly detection for astronomy applications. For supernova (SN) light curves,  \citet{ishida2019active, pruzhinskaya2019anomaly, aleo2020most,2020arXiv200906760M} used isolation forests and active anomaly discovery on archival datasets. Convolutional autoencoders have recently been used to search for anomalies and glitches in gravitational wave time series \citep{2021arXiv210307688M}. In the broader machine learning literature, there has been increasing interest in active anomaly detection \citep{chalapathy2019deep}. Typically these works focus on long and well-sampled, single channel time series with anomalous periods of activities  (e.g., \citealt{zhang2019deep, 2019arXiv190104997L}), although some recent work has focused on multivariate series \citep{zhao2020multivariate}.  There has been limited focus on anomaly detection in irregularly sampled, aperiodic, multivariate time series.

In this paper, we focus on out-of-distribution anomalies, which appear distinct from all other known transients in some feature space. By taking a completely data-driven approach to anomaly searches, our algorithm is agnostic to physics and therefore sensitive to entirely unexpected phenomena. Our anomaly detection pipeline is based on a variationl recurrent autoencoder neural network (VRAENN) with no physical priors, and we search the learned encoded space for out-of-distribution events. The paper is organized as follows. In \S2 we review the \plasticc\ dataset used for training and anomaly detection, and the breakdown of SN-like transients used in this study. In \S3 we present our anomaly detection pipeline and the VRAENN architecture. We discuss our results in \S4 and conclude in \S5. 

\begin{figure*}
  \centering
  \includegraphics[width=\textwidth]{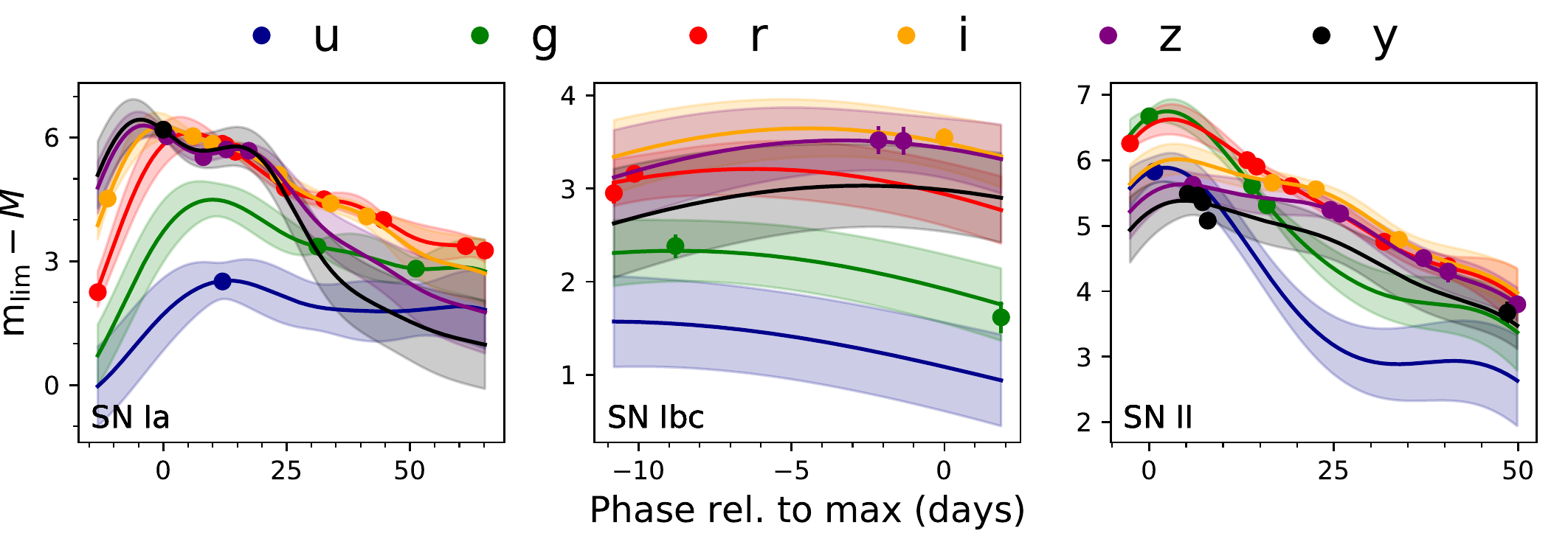}
  \caption{Sample $grizY$ light curves of majority SN classes (Type Ia, Ibc, II). Bold lines represent the 2D Gaussian Process mean function, with shaded regions representing 68\% confidence intervals. Even when entirely missing one or more bands, our method is able to produce reasonable interpolated light curves. Note that the y-axis, the magnitudes used to train the Gaussian Process, is designed such that the light curve tends to zero as the light curve approaches the survey magnitude limit. \label{fig:lc}}
\end{figure*}

\section{Dataset and Preprocessing}

In this study we use PLAsTiCC ({\it Photometric LSST Astronomical Time-Series Classification Challenge}), a simulation of three years of Rubin Observatory data that includes over 3.5 million transient events from eighteen unique physical classes, extending to a redshift of $z\approx1.5$ \citep{allam2018photometric,kessler2019models}. Each event is observed across six broadband filters ($ugrizY$) following the LSST observing strategy at the time the simulation was produced.  Sample light curves are shown in Figure~\ref{fig:lc}). Along with the light curves, PLAsTiCC provides metadata including the redshift, Milky Way reddening, physical parameters used to generate the model, and a realistic photometric redshift estimate (see \citealt{kessler2019models} for details). For this observing strategy, each event is observed every few days (in any filter) and roughly once a week in the same filter.

The PLAsTiCC dataset was originally created for a public data science (Kaggle) competition\footnote{\url{https://www.kaggle.com/c/PLAsTiCC-2018}} to classify transients. We re-purpose this dataset as a training set for anomaly detection in an LSST-like data stream. Here, anomalous events will be determined by the metadata (i.e., if the event comes from a rare astrophysical origin). We remove classes observable only within the Milky Way (e.g., variable stars) and SNe with fewer than three \textit{detections} in any filter within 300 days of peak brightness. This cut is not necessary, as our algorithm can take light curves of any length; however, light curves with fewer points are very unlikely to be selected for detailed follow-up in reality. Additionally, we only utilize transients from the Wide-Fast-Deep (WFD) survey and remove events within the Deep Drilling Fields. In total, our data set contains \Ntrans\ extragalactic light curves from thirteen classes: 

\begin{itemize}
    \item \textbf{Normal Type Ia SNe} arise from the thermonuclear explosions of carbon-oxygen white dwarfs. The models were generated using the standard SALT-II light curve models \citep{guy2007salt2}, conditioned on $\approx 500$ light curves from the Joint Lightcurve Analysis \citep{betoule2014improved}. Type Ia SNe represent \peria\ percent of our dataset. We consider Type Ia SNe part of the \textbf{majority} classes.
    
    \item \textbf{Type II SNe\footnote{\label{footnote-label} In the original version, PLAsTiCC grouped normal Type II SNe and Type IIn SNe into a single class. We separate these classes due to their distinct physical origins and unique light curves.}} are the explosions of massive stars that have retained their hydrogen envelopes. They are often characterized by long plateaus in their optical light curves. The models were generated from SED templates \citep{kessler2010supernova,anderson2014characterizing,galbany2016ubvriz,sako2018data}. Type II SNe make up \perii\ percent of our dataset. We consider Type II SNe part of the \textbf{majority} classes.
    
    \item \textbf{Type Ibc SNe} are core-collapse SNe of stars with stripped hydrogen (Ib) and helium (Ic) envelopes. The models were generated using a combination of MOSFiT \citep{villar2017theoretical,guillochon2018mosfit} and SED templates \citep{kessler2010supernova}. Type Ibc SNe make up \peribc\ percent of our sample. We consider Type Ibc SNe part of the \textbf{majority} classes.
    
    \item \textbf{Type I Superluminous SNe} (SLSNe) are luminous, hydrogen-free events thought to be powered by rapidly spinning, highly magnetized neutron stars. SLSN models were produced using MOSFiT \citep{guillochon2018mosfit,nicholl2017magnetar,villar2018superluminous}. They make up \perslsn\ percent of our sample. We consider SLSNe to be members of the \textbf{minority} classes.

    \item \textbf{Type Iax} are irregular Type Ia SNe with typically lower luminosities and lower velocities compared to normal Type Ia SNe \citep{li2003sn}. The models were generated using available data in the Open Supernova Catalog \citep{guillochon2017open}. Type Iax SNe make up \periax\ percent of our dataset. We consider Type Iax SNe to be members of the \textbf{minority} classes.

    \item \textbf{Type IIn SNe\textsuperscript{\ref{footnote-label}}} are core-collapse SNe mainly powered by the interaction of the SN ejecta with circumstellar material (CSM). The models were generated using MOSFiT \citep{villar2017theoretical,guillochon2018mosfit,jiang2020extended}. Type IIn SNe make up \periin\ percent of our sample. We consider Type IIn SNe to be members of the \textbf{minority} classes.

    \item \textbf{Type Ia-91bg} are fainter, faster and redder Type Ia SNe that make up $\approx 20$\% of the volumetric Type Ia sample \citep{filippenko1992subluminous,graur2017loss}, and $\approx 3$\% of the observational sample \citep{li2011nearby}. The model light curves are based on the spectral energy distribution (SED) templates from \citet{nugent2002k}. Type Ia-91bg SNe make up \perbg\ percent of our sample. We consider Type Ia-91bg SNe to be members of the \textbf{minority} classes.

    \item \textbf{Tidal Disruption Events} (TDEs) result from the tidal disruption of stars by supermassive black holes (SMBH; \citealt{rees1988tidal}). TDE models were generated using MOSFiT \citep{guillochon2018mosfit,mockler2019weighing} and make up \pertde\ percent of our sample. We consider TDEs to be members of the \textbf{minority} classes.

    \item \textbf{Ca-rich Transients} (CARTs) are intermediate luminosity transients whose spectra appear rich in calcium \citep{kasliwal2012calcium}. CARTs are modeled using MOSFiT, assuming they are powered by the radioactive decay of $^{56}$Ni. We note that  this is the same model used to generate Type Ibc SNe, but with a distinct parameter space. CARTs make up \percart\ percent of our sample. We consider CARTs to be members of the \textbf{minority} classes.

    \item \textbf{Intermediate Luminosity Optical Transients} (ILOTs) are transients that are brighter than novae but less luminous than SNe \citep{kasliwal2012systematically}. In this case, we assume that ILOTs arise from CSM interaction with low-energy eruptions (or explosions) of massive stars. ILOTs have been modelled using MOSFiT \citep{guillochon2017open,villar2017theoretical,jiang2020extended} and represent \perilot\ percent of our sample. We consider ILOTs to be members of the \textbf{minority} classes.

    \item \textbf{Pair Instability Supernovae} (PISNe) are the explosions of low-metallicity massive stars ($M_\mathrm{ZAMS}\sim130-260$ M$_\odot$) that reach core temperatures high enough to form electron-positron pairs \citep{kasen2011pair}. Compared to normal core-collapse SNe, PISNe have high kinetic energies and larger ejecta masses. PISNe are modelled using MOSFiT assuming that they are powered by the radioactive decay of $^{56}$Ni. PISNe make up \perpisn\ percent of our sample. We consider PISNe to be members of the \textbf{minority} classes.

    \item \textbf{Kilonovae} (KNe) arise from the formation of radioactive rapid neutron capture elements in binary neutron star (and potentially neutron star - black hole) mergers. The models are based on theoretical calculations \citep{kasen2017origin}. There are only two KNe in our sample. KNe are dim and short-lived, making them nearly impossible for LSST to discover in the WFD strategy explored here. We consider KNe to be members of the \textbf{minority} classes.
    
    \item \textbf{Active Galactic Nuclei} (AGN) refers generally to galaxies with active SMBHs from accreting gas. AGN have a wide range of observed behavior, although typically they vary on timescales of weeks to years at the $\lesssim 10$\% level. AGN variability is modeled using a damped random walk as described in \cite{macleod2011quasar}. AGN make up \peragn\ percent of our dataset. We note that AGN are the only class that is likely \textit{not} representative of the true LSST datastream; AGN will be much more numerous, with likely millions in the complete sample. Many bright AGN (which are those represented in \plasticc) will be identified within the first year of LSST; however, transient bright flares will be of interest to the community \citep{graham2017understanding}. Because AGN are distinct from SN-like transients, we consider their effects separately in an Appendix.

\end{itemize}

\begin{figure}
  \centering
  \includegraphics[width=0.48\textwidth]{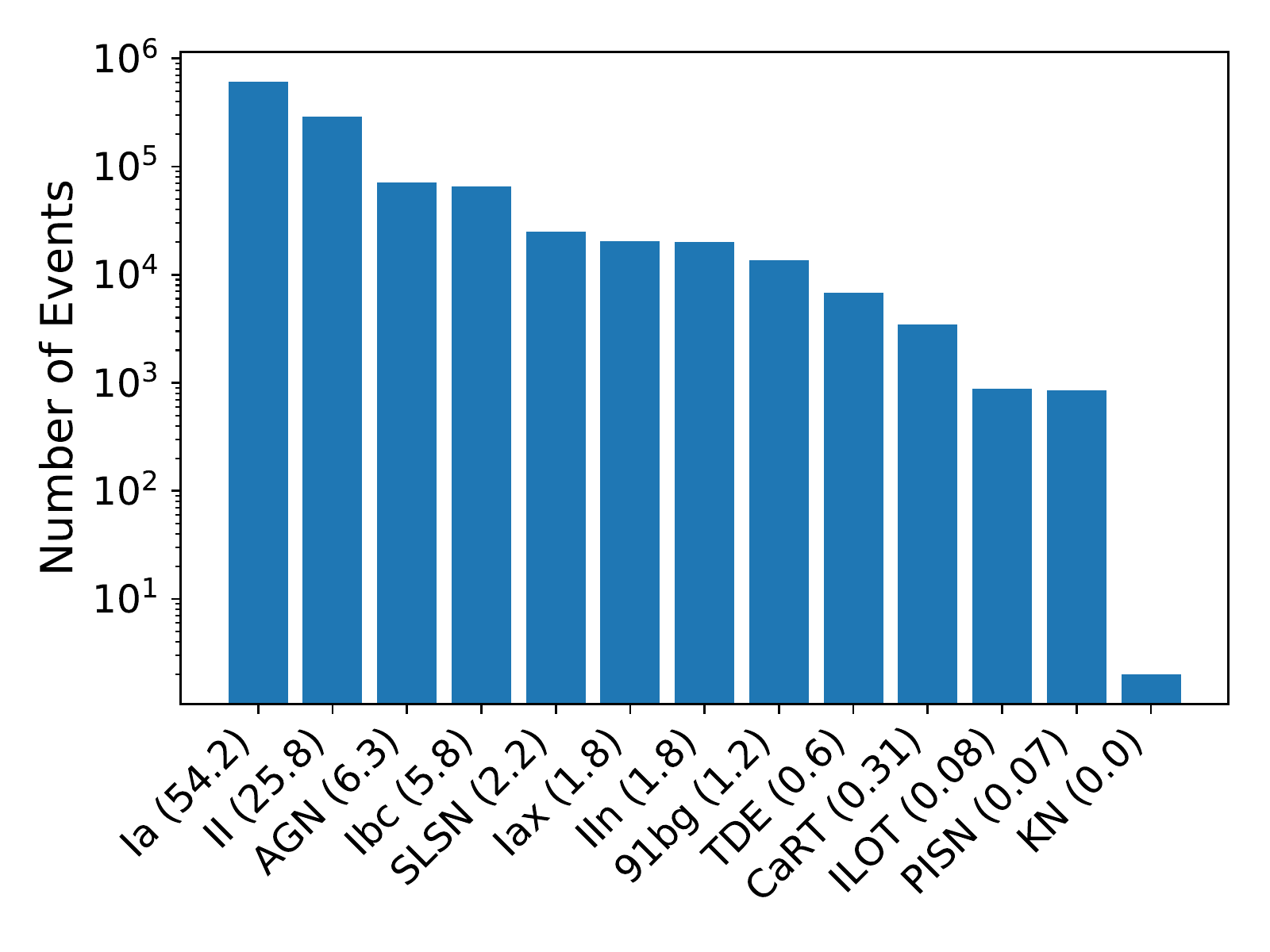}
  \caption{Breakdown of the various transient classes used in this study. Note that AGN are likely highly underrepresented compared to the true breakdown in the LSST datastream; however, AGN are exceptional in their light curve properties and may be identified early in the survey. Parenthetical numbers associated with each class represent the percentage breakdown. There are a total of two kilonovae (KNe) in our sample. \label{fig:hist}}
\end{figure}

The breakdown of the various classes is shown in Figure~\ref{fig:hist}. The observational rates of each class are a combination of the volumetric rates (intrinsic rarity) and observational effects (e.g., luminosity function, duration).  The  breakdown used here is designed to match what is empirically expected from a wide-field survey such as LSST (see e.g., \citealt{perley2020zwicky,villar2020superraenn}). We note that this LSST cadence simulation detects just two KNe that pass our cuts in the simulated observations, highlighting the need for target-of-opportunity observations to better capture these rare events. 

We define Type Ia SNe, Type II SNe and Type Ibc SNe together as the majority classes because together they make up the bulk of the dataset ($\approx 86\%$). We consider all other classes (excluding AGN)  to be minority classes, with fraction of  $\lesssim 2$\% of the sample.

We pre-process the data as follows. Following \citet{villar2020superraenn}, we scale the light curve magnitudes such that zero corresponds to the magnitude limit. This is to aid the Gaussian process interpolation (which will tend towards zero before and after the supernova). We correct the light curve of each event for time dilation and convert to absolute magnitudes based on the provided photo-$z$ values (see \citealt{villar2020superraenn} for details). The photo-$z$ values are based on host galaxy association, and use a color-matched nearest neighbor method presented in \citet{graham2018}. This method is trained on a realistic sample of galaxies that would have spectroscopic redshifts  available. About 17\% of the redshift estimates are outliers, defined as $|z_\mathrm{true}-z_\mathrm{phot}|/(1+z_\mathrm{phot})>3\sigma_\mathrm{IQR}$ by \citealt{kessler2019models}, where $\sigma_\mathrm{IQR}$ is the typical error for galaxies near $z_\mathrm{true}$. We make no cuts on redshift uncertainty, and instead  we account for redshift uncertainties via a simple Monte Carlo method discussed in \S\ref{sec:vraenn}. 

We additionally correct the light curves for Galactic reddening, as outlined in \citet{villar2020superraenn}. Finally, we temporally shift each event such that the observed time of peak brightness (in any filter) is considered $t=0$. This means that as new data are taken during an event's rise, the time of peak luminosity will continue to shift until the true peak luminosity has been observed or the event dims. For example, if a transient first peaks in $g$-band 10 days post explosion and then reaches a brighter luminosity in $r$-band at 15 days post explosion, the peak used by the neural network will be at 15 days post explosion; this is independent of the bolometric peak luminosity. We find that this prescription of phase (versus, for example, time since first detection) leads to better performance in the autoencoder.

\subsection{Interpolation using Gaussian Processes}

The PLAsTiCC light curves are irregularly sampled across time and filters, with no more than one filter observed at any given time. For the architecture discussed below, we require a flux and error estimate for each filter at every observation time.  To produce this information, we use a 2D Gaussian Process (GP) to interpolate the light curve over time and filter, with a multivariate Gaussian kernel described by:
\begin{multline}
\kappa(t_i, t_j, f_i, f_j; \sigma, l_{t}l_{f})= \\
\sigma^2 \times \exp\Big[-\frac{(t_i-t_j)^2}{2l_t^2}\Big]
\times\exp\Big[-\frac{d(f_i, f_j)^2}{2l_f^2}\Big]
\end{multline}
where $f$ represents the six ($ugrizY$) filters; $l_t$ and $l_f$ are characteristic correlation length scales in time and filter, respectively; and $d(f_i, f_j)$ is the Wasserstein-1 distance between each filter's normalized throughput which we optimize to each specific light curve. This choice in the distance metric loosely measures the similarity between two distributions. Mathematically, we treat each filter as a density function in wavelength. This distance metric is minimized when filters overlap and simplifies to difference between central wavelengths in the limit of infinitely narrow passbands. 

The GP interpolation both accounts for and produces error estimates for each flux measurement. Flux uncertainties are utilized in both the encoding method and anomaly scoring steps of our algorithm (see \S\ref{sec:vraenn}) in order to make our algorithm robust to low-confidence outliers. Finally, we note that during testing we implemented a similar (though less physically motivated) interpolation scheme in \citet{villar2020superraenn} and found that the interpolation methods led to visually similar light curves. 

We implement the GP pre-processing using {\tt sklearn}, optimizing each light curve independently via the {\tt minimize} function from {\tt scipy}, which uses the Broyden–Fletcher–Goldfarb–Shanno optimization algorithm \citep{fletcher1986}. Sample light curves are shown in Figure~\ref{fig:lc}. Even in cases of poorly sampled light curves, or light curves in which a band is completely missing, the GP produces reasonable flux and error estimates across all filters.

We note that our GP utilizes the complete light curve for interpolation. In reality, only the light curve up to the most recent observation will be available in real time. One may be concerned that because our GP has been conditioned on the entire light curve, it has more information than what will be available for real-time usage. This likely has no effect on our results, as each observation heavily anchors the GP prediction (see Figure~\ref{fig:lc}) and the learned GP kernel sizes are similar to our priors (see e.g., \citealt{villar2020superraenn} for typical SN values).

\section{VRAENN: Architecture, Training and Anomaly Detection}\label{sec:vraenn}

\begin{figure*}
  \centering
  \includegraphics[trim={0 0cm 0 0cm},clip,width=\textwidth]{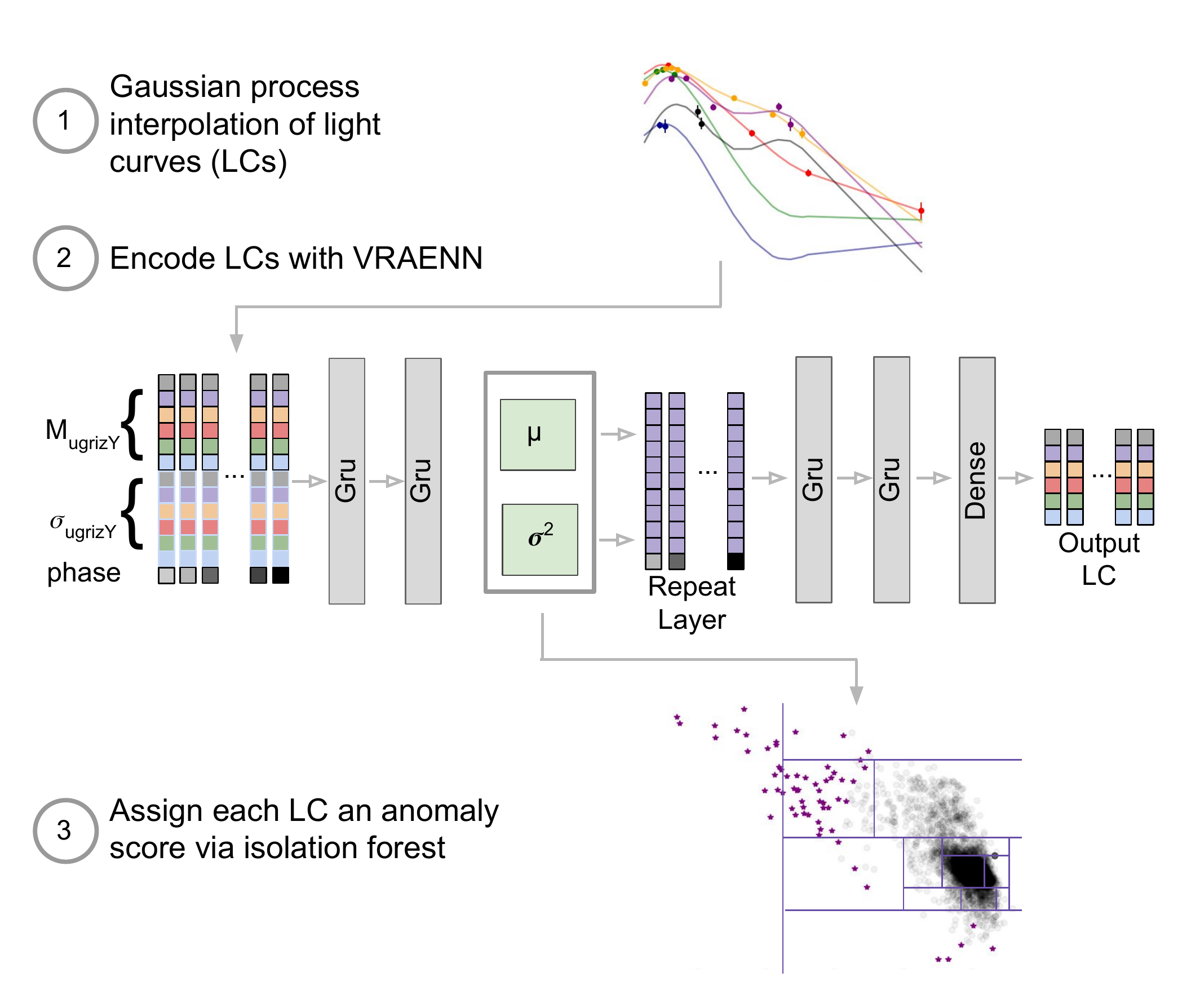}
  \caption{Summary of our anomaly detection pipeline. (1) We interpolate each $ugrizY$ light curve using a 2D Gaussian process. (2) We train the variational recurrent autoencoder in an unsupervised manner. Light curves are represented as a time series, in which each epoch is represented by 11 features: 6 flux values, 6 estimated error values, and one time. This time series (consisting of $N$ points) is encoded. The encoding layer is repeated $N$ times, each time appended with the associated time value. We use this network to encode each light curve. (3) We use an isolation forest to assign an anomaly score for each light curve using the encoded vectors. The isolation forest is represented here in one subspace of the encoded space, with anomalous events highlighted as purple stars. \label{fig:arch}}
\end{figure*}

Following pre-processing of the training set, our anomaly detection pipeline includes two steps. First, we learn an encoded form of each light curve by training a variational recurrent autoencoder neural network (VRAENN) on the full dataset. Then, we use an isolation forest to rank each light curve's encoded form by an anomaly score. We utilize a simple Monte Carlo to estimate the uncertainty on this score. In this section, we describe the VRAENN architecture, the training process, the isolation forest and our error estimation method. Our code, along with all chosen hyperparameters of our model, is available via Github\footnote{\url{https://github.com/villrv/vraenn}}

Our novel VRAENN architecture is well-suited to the problem of searching for unknown, anomalous transients. While more traditional feature extraction via model-fitting (e.g., \citealt{villar2018,hosseinzadeh2020photometric}) or using pre-defined quantities \citep{boone2019avocado} has been successful for classification, dozens of features are required. We want to avoid searching for anomalous events in high-dimensional data, in which distance metrics are more challenging to meaningfully define \citep{liu2012isolation}. Additionally, the VRAENN architecture is built without any physical models, meaning that it is sensitive to new and unexpected physical processes, which are observationally distinct from known transients. 

The architecture specifically used here has three additional benefits: (i) The ability to handle unevenly sampled light curves across filters and time using recurrent neurons; (ii) the ability to produce extrapolated and interpolated light curves; and (iii) an insensitivity to noisy data and spurious outliers due to the variational architecture and error estimation. 

The VRAENN architecture, based on the model presented in \citet{villar2020superraenn}, uses recurrent neurons to read in the GP light curve and estimated errors, and encodes this light curve as a vector. This is achieved by encoding the light curve into a series of smaller matrices until the information reaches a small bottleneck layer of size $1\times10$. This is an unoptimized choice of size, although \citet{villar2020superraenn} found a similar result after a hyperparameter search of a similar architecture. This layer is known as the encoded layer, and the layers preceding it are known as the ``encoder''. Each of these layers uses gated recurrent unit (GRU) neurons with a combination of hyperbolic tangent, sigmoid, relu and linear activation functions. The GRU is memory-efficient version of the Long short-term memory, the standard for RNNs \citep{2014arXiv1406.1078C}.

Rather than using a static vector as a bottlebeck, our VRAENN learns a \textit{distribution} of encoding vectors. In particular, the encoding layer consists of two vectors: one that represents the mean of a multivariate Gaussian (MVG) and one that represents its diagonal covariance matrix. For each SN passed into the VRAENN, we randomly select an encoding from the MVG defined by this \textit{learned} mean and variance. The ``variational'' aspect of our architecture refers to this process of learning a distribution of encodings rather than a singular encoding. This is helpful in generating a smooth encoding space, in which most events will cluster, allowing us to more easily pick out anomalous events.

\begin{figure*}
  \centering
  \includegraphics[width=\textwidth]{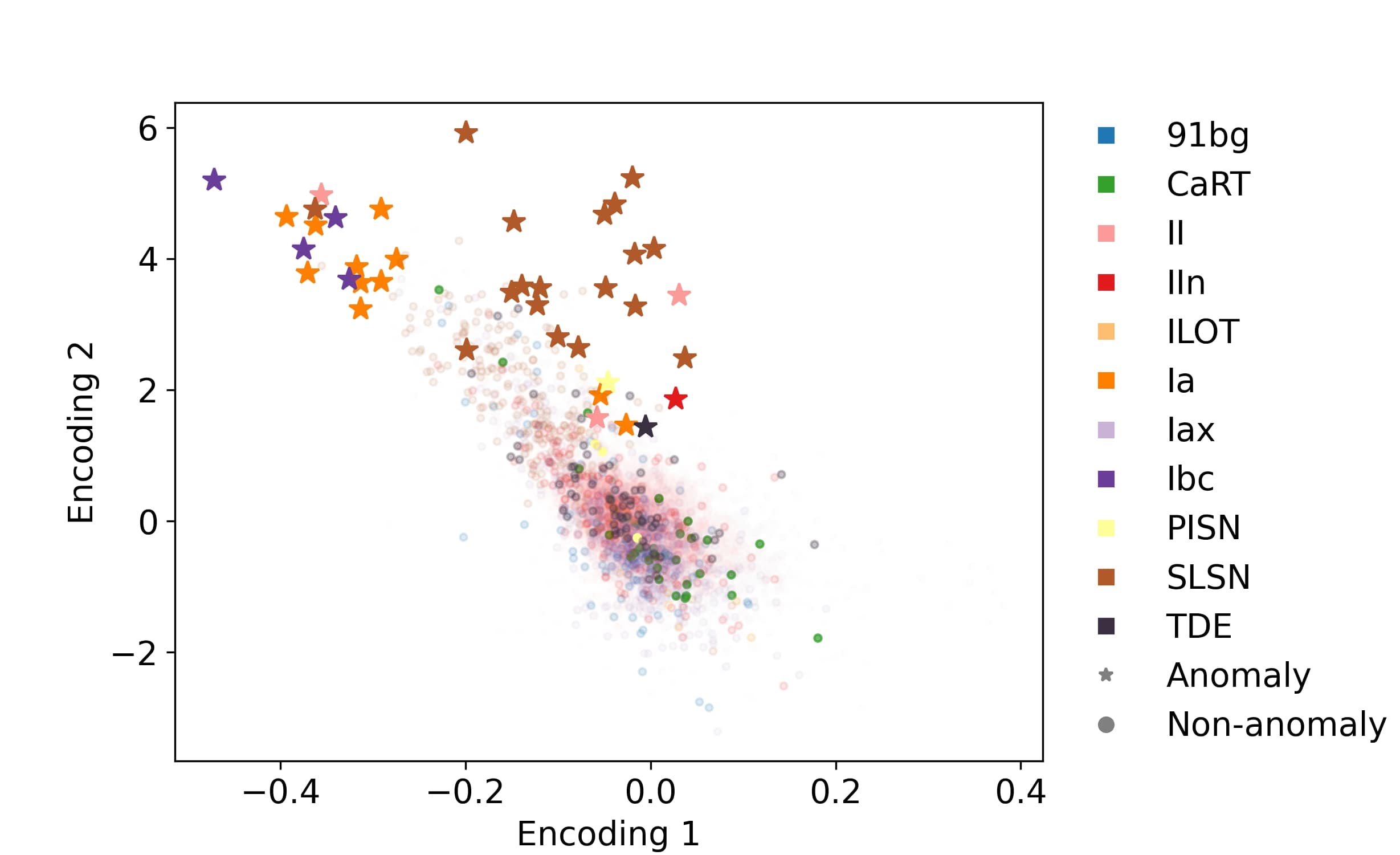}
  \caption{Scatter plot of representative encoding space for various classes. Events with high anomaly scores are shown as stars. Most events cluster near the origin, while more anomalous classes are seen as clouds outside the main distribution. The anomalous Type Ia, Ibc and II SNe are those with incorrect photo-$z$ estimates. \label{fig:encspace}}
\end{figure*}

Before being passed into the second half of the autoencoder (the ``decoder''), the encoded layer is repeated $N$ times, with each time appended as a phase relative to maximum light. This can be thought of as evaluating a SN model at specific times, where the model is specified by 10 free parameters, with an 11th parameter specifying the time. The unique repeat layer of our architecture allows us to evaluate the light curve at times not included in the real data; i.e., interpolating and extrapolating the light curve if desired (although this feature is not used in this work).

The decoder then produces the light curves at the $N$ times specified in the repeat layers. As with the encoder, the decoder uses GRU neurons with hyperbolic tangent activation functions. Figure~\ref{fig:arch} illustrates our full pipeline, including a schematic of the neural network architecture used. The VRAENN is optimized using a loss function combining the log of the weighed mean squared error and the standard Kullback–Leibler (KL) divergence (which measured how well our MVG represents our latent variable space):

\begin{multline}
    \LikeL=
\log\sum_{i=1}^N\frac{\Big[F_{i,\mathrm{True}}(t,f) - F_{i,\mathrm{Predicted}}(t,f)\Big]^2}{N}\\
+\sum_{i=1}^N -0.5\Big(1+\log(\sigma_i^2) - \mu_i^2 - \sigma_i^2\Big)
\end{multline}

We minimize the loss function using the Adam optimizer \citep{2014arXiv1412.6980K} with standard learning parameters, $\alpha=10^{-4}$, $\beta_1=0.9$, $\beta_2=0.999$ using Keras \citep{chollet2015keras} with a TensorFlow backend \citep{tensorflow}. We train our VRAENN on 1\% of the sample (12,159 events), reflecting (for example) the small dataset that will be available within months of LSST coming online.

A sampled subspace of our encoding vectors is shown in Figure ~\ref{fig:encspace}. The majority of events cluster near zero, with anomalous events (like SLSNe) forming a cloud outside of the main distribution.

\subsection{Scoring Anomalies with an Isolation Forest}

Once our VRAENN is trained, we can encode any \plasticc\ light curve, partial or complete, as a $1\times10$ vector. We then pass these encodings into an isolation forest \citep{liu2012isolation}. The isolation forest works by generating a series of decision trees over a random subset of attributes. Each tree recursively splits the set. Out-of-distribution anomalous events will be isolated with very few splits, while an average event will cluster with similar events, even after many splits. The number of splits is inversely related to an anomaly score. For the sake of interpretability, we then convert this raw score to a percentile. We use {\tt sklearn} to implement the isolation forest, using 1000 base estimators. 

We identify several sources of possible error in the anomaly score:

\begin{itemize}
    \item Flux uncertainty due to Poisson noise. This is provided by \plasticc\ as a standard deviation for each flux measurement.
    \item Uncertainty in the flux \textit{estimates} from the GP, which is also estimated by the GP.
    \item Photometric redshift error, reported as a standard deviation. This affects the entirety of the light curve as both an overall multiplicative flux term and a time dilation adjustment.
    \item Model uncertainty from the neural network converting the light curve into an encoded vector.
\end{itemize}
We account for the first three using a simple Monte Carlo method. For each transient, we generate 10 light curves that have photo-$z$ and flux values drawn from Gaussian distributions described by the reported mean and errors. We do not account for noise from the neural network itself, which could be accomplished via an ensemble of networks. We find that, in general, the error estimated from this method is sufficient to eliminate anomalous events arising from poor data quality and incorrect photo-$z$ estimates by making a cut on the anomaly score uncertainty. 

\begin{figure}[t!]
  \centering
  \includegraphics[width=0.45\textwidth]{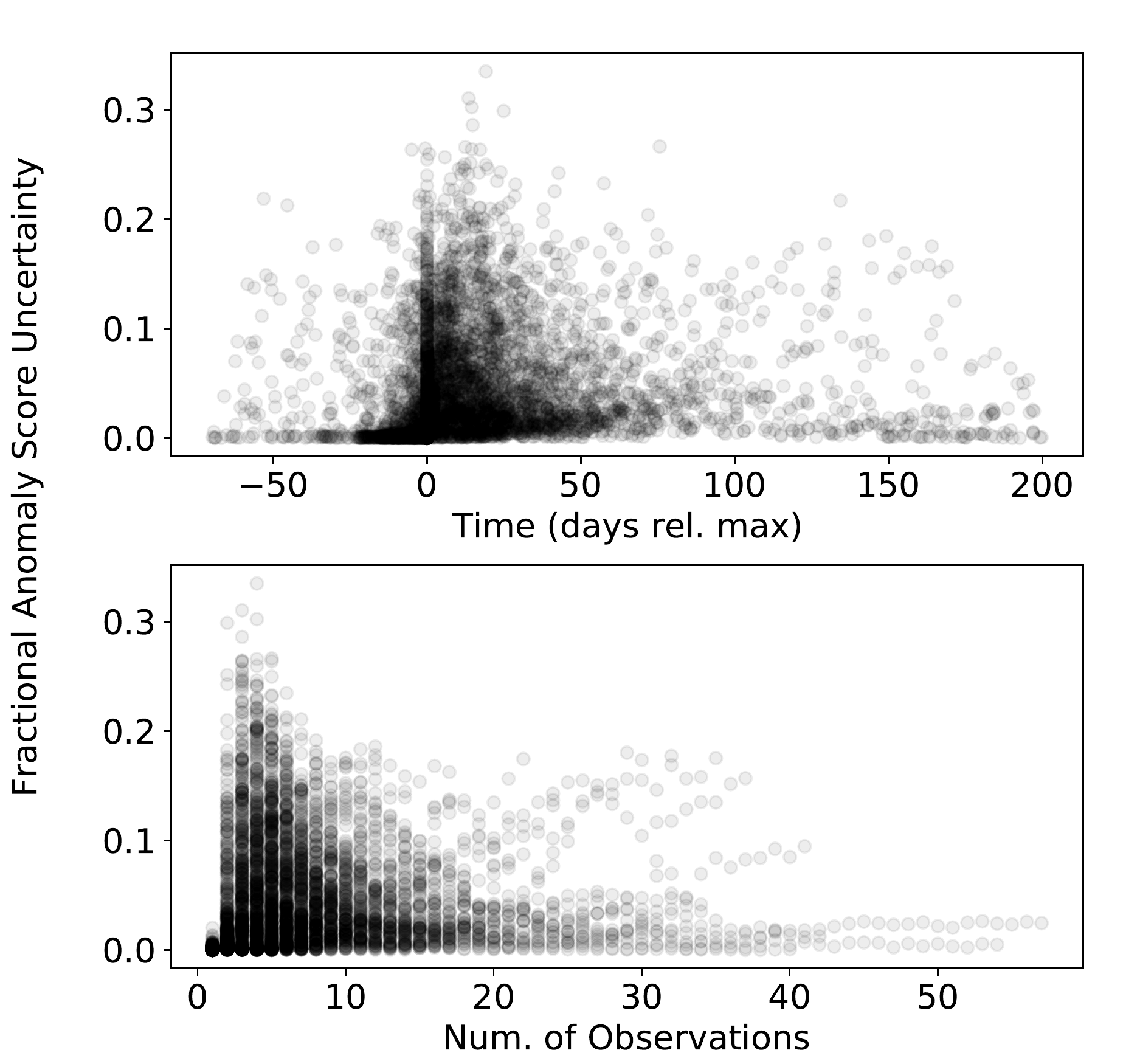}
  \caption{Time evolution of the relative anomaly score error for a representative subsample of our test set. In the upper panel, relative errors are low before peak luminosity and subsequently rise; however, the lower panel reveals that the relative error drops with increased number of data points. \label{fig:anomerr}}
\end{figure}

We show how the anomaly score uncertainty changes as a function of both time and number of data points in Figure~\ref{fig:anomerr}. We first examine how the error changes as a function of number of observations. The error grows until the light curve reaches $\approx3$ observations and then decreases. The error is initially small because the autoencoder, with limited information, returns the mean encoding vector from the training set (i.e., only one number is needed to encode a single observation). These light curves are not flagged as anomalous. However, once the light curve reaches a sufficient number of data points, the fractional uncertainty follows the expected trend of decreasing with increasing number of observations. When translated into fractional error versus time, the fractional error rises near peak (where most light curves have few data points) and declines at late times.

\section{Results and Discussion}

Our anomaly classification algorithm generates a ranked list of anomalous events given the full dataset. We are interested in which events the anomaly detection pipeline ranks highly, but we are also interested in understanding \textit{when} and \textit{why} an event becomes anomalous. 

\begin{figure}[t!]
  \centering
  \includegraphics[width=0.5\textwidth]{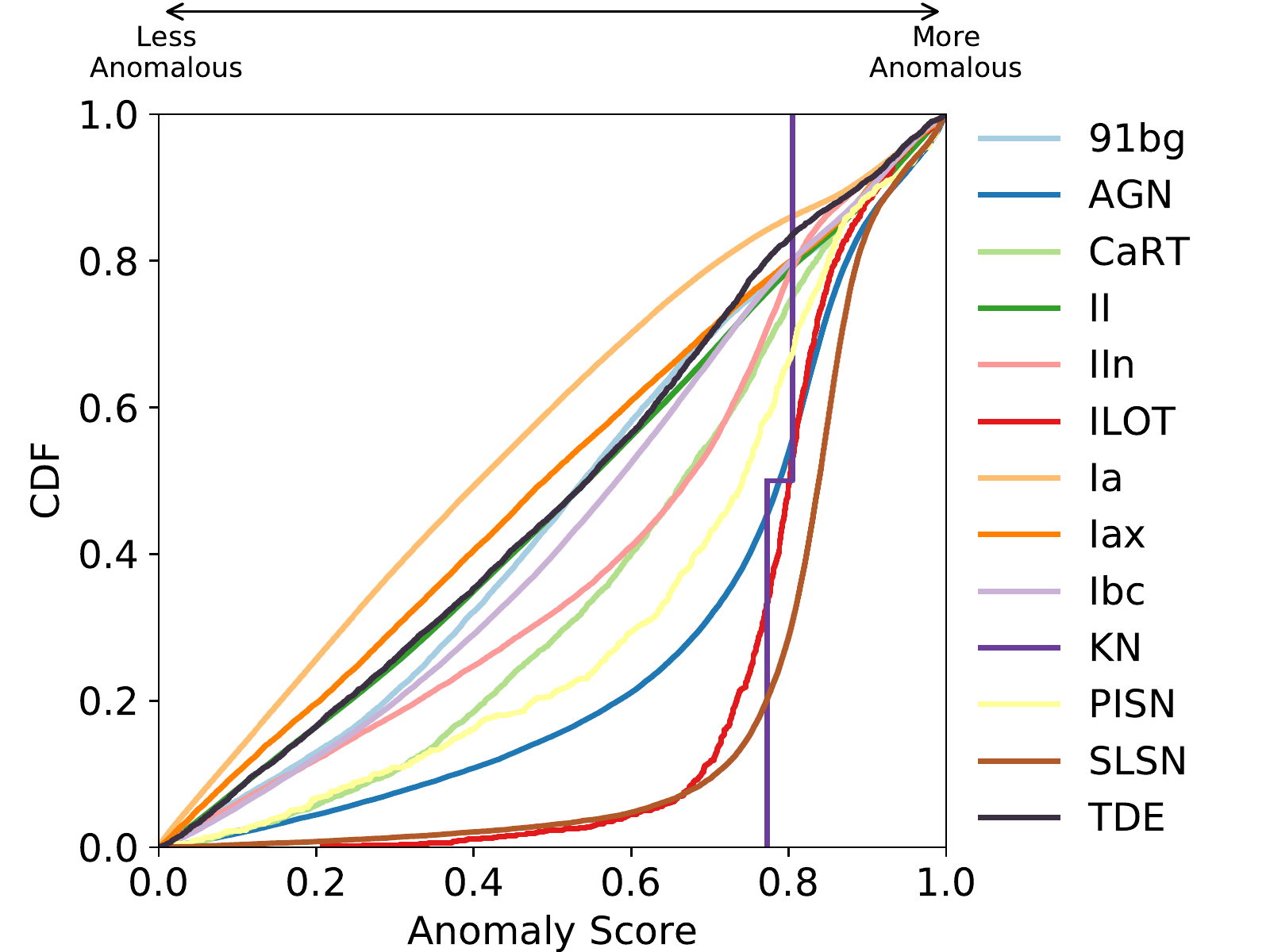}
  \caption{ CDF of anomaly score for various classes of transients.  PISNe, SLSNe, ILOTs, KNe and AGN are clustered towards the high end of anomaly score. \label{fig:cdf}}
\end{figure}

We first explore the anomaly scores of the full test set. In Figure~\ref{fig:cdf}, we show the cumulative distribution function (CDF) for each class in our set. Even for events that would not be labelled as anomalous by our pipeline, the anomaly score distribution matches expectations. Type Ia-like SNe (including 91bg, Iax) are the least anomalous on average. Type Ibc SNe and TDEs are largely distributed evenly across scores, with a slight tail at the upper end. While Type II SNe are, on average, more anomalous, only a small fraction have high anomaly scores. CARTs and Type IIn SNe are also more anomalous on average. PISNe, SLSNe, ILOTs and AGN most drastically cluster at the high end of anomaly scores, indicating that they are the most likely to be classified as anomalies with our algorithm.  Only two KNe were in our sample; both had moderately high anomaly scores around the 80th percentile. Reassuringly, the events which prefer higher anomaly scores do not cluster in any obvious section of observational phase space (e.g., luminosity or duration), implying that our VRAENN has picked up on more fundamental features.

\begin{figure*}
  \centering
  \includegraphics[width=\textwidth]{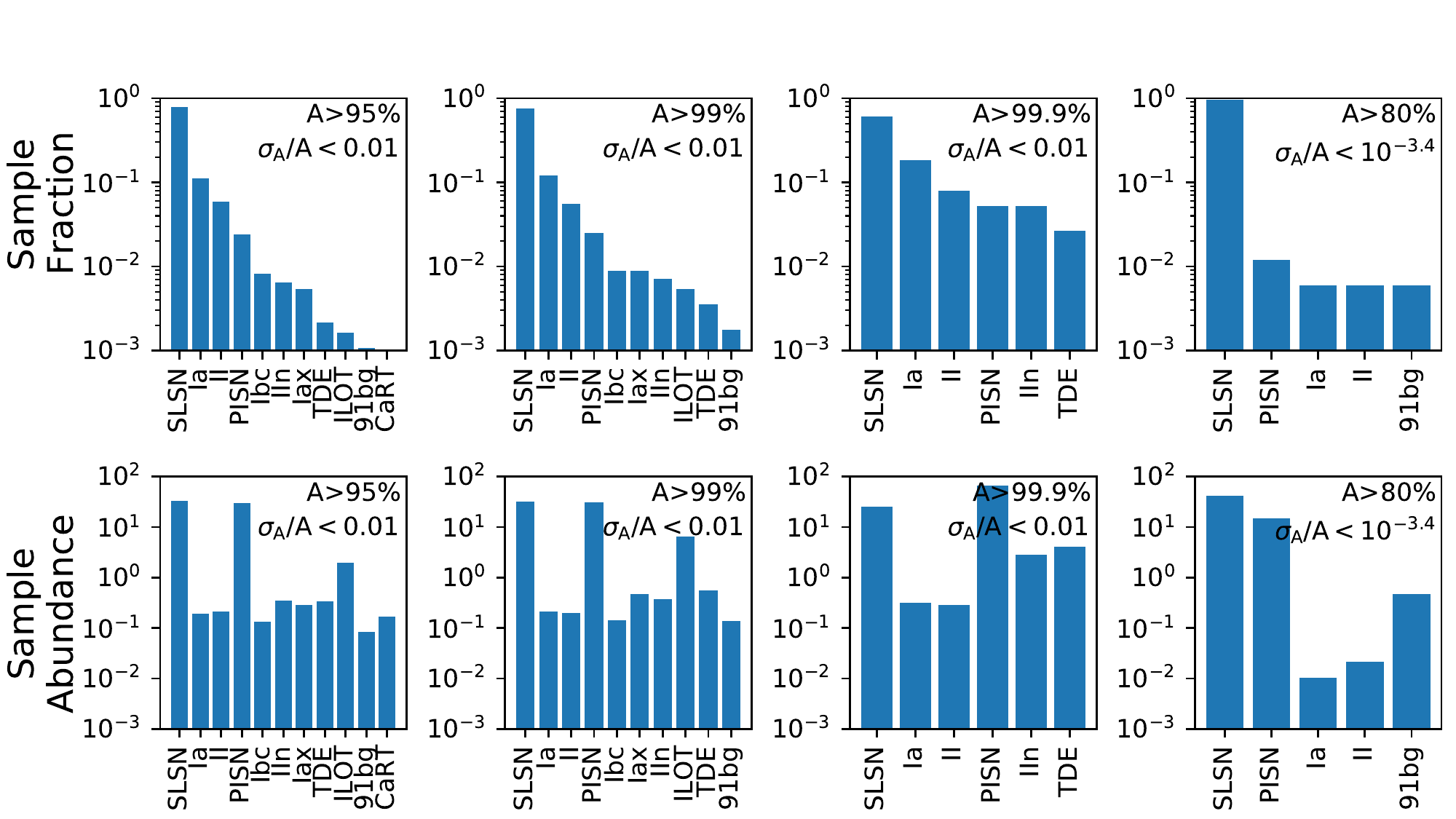}
  \caption{\textit{Top row:} Breakdown of anomalous events, given the cuts listed in each panels. The top number represents the anomaly percentile threshold, while $\sigma_\mathrm{A}/\mathrm{A}$ refers to the anomaly score fractional error threshold. The right-most panel shows optimized cuts on the anomaly score and uncertainty to minimize contamination from majority classes. SLSNe overwhelmingly make up the majority of our anomaly sample. \textit{Bottom row:} Relative abundance of anomaly samples shown in top rows. Each panel shows the ``abundance'' (ratio between the number of ``anomalous'' events in each class and the expected number based on input fraction. The minority classes are heavily over-represented, while majority classes are underrepresented.  \label{fig:histdist}}
\end{figure*}

In practice, we will be limited to a small fraction ($\sim 0.1$\%)  of the LSST transients for follow-up. We therefore investigate several thresholds on the anomaly scores and uncertainties to search for anomalous events. The results presented here are summarized as histograms of anomaly sample breakdown and abundances in Figure~\ref{fig:histdist}. In short, our anomaly sample is \textit{pure}, with $\lesssim 10$\% contamination from majority (Type Ia, Ibc, and II SNe) classes. The anomaly sample significantly over-represents minority classes.  

Many of the non-anomalous transients (e.g., Type Ia SNe) with high anomaly scores have fractional uncertainties on the anomaly scores of $\sigma_\mathrm{A}/\mathrm{A}\gtrsim 0.1$ (due to large uncertainties on photo-$z$ estimates). We first make a strict cut on the full sample, looking only at events in the top 90, 95, and 99 percentile of anomaly scores, and with $\sigma_\mathrm{A}/\mathrm{A}<0.01$. Within these cuts, we find that SLSNe make up a majority of the remaining sample. For the 95 percentile cutoff, the anomalous SLSNe that pass our cut make up $\approx5$\% of the original sample. In contrast, just 0.03\% of the original input sample for Type Ia and Type II SNe remain in our 95 percentile anomaly score cutoff. In other words, our 95 percentile cutoff removed 95\% of the SLSN sample, but removed 99.97\% of the Type Ia and Type II SNe samples (leading to SLSNe being over-represented in the anomalous sample by a factor of about 4.5). Similarly, PISNe make up a small fraction ($\approx1$\%) of our anomalous sample, but we retain $\approx 5$\% of the original sample in the 95 percentile cutoff. The results are similar for higher percentiles, with an even larger bias towards the more anomalous classes. This is highlighted as an abundance measurement in Figure~\ref{fig:histdist}.

We then explore a higher cutoff in anomaly score fractional error, keeping events with $\sigma_\mathrm{A}/\mathrm{A}<4\times 10^{-4}$ (a threshold chosen by hand to maximize the ratio of rare events, shown in the right-most column of Figure~\ref{fig:histdist}). Even without cuts on anomaly score itself, SLSNe dominate the sample, followed by PISNe and Type Ia-like SNe. 

Noting that stringent cuts in anomaly score uncertainty lead to pure samples of minority classes, we investigate if incorrect photo-$z$ estimates are responsible for false positive detection of anomalous events in majority classes. We find that removing events with $|z_\mathrm{phot}-z_\mathrm{true}|/z_\mathrm{true}>2$ does improve the purity of anomalous samples for cuts with $A>99$\% and $A>99.9$\%, although has a lesser effect for $A>95$\% cut. For all cuts, we find that removing events with incorrect photo-$z$ estimates drastically decrease the number of Type Ia SNe which pass our anomaly thresholds. For example,  in the $A>95$\% sample, the sample fraction of Type Ia SNe drops from $\approx10$\% to $\approx3$\%. Reducing the number of outlier photo-$z$ estimates in LSST will likely substantially improve the purity of our anomaly sample. 

We investigate why some Type Ia SNe are within the most anomalous events even with stringent cuts on anomaly score and uncertainty. All of the events which pass our 99\% cutoff for anomaly score have catastrophically incorrect photo-$z$ values. Specifically, we find that these events are typically injected at relatively low redshifts ($z_\mathrm{true} \lesssim 0.05$) yet have reported photometric redshifts $z_\mathrm{phot} \approx 2$ with small reported uncertainties---again highlighting the need for reliable photo-$z$ estimates.

Our anomaly detection algorithm is biased towards bright events, which begs the question: Is our algorithm making a trivial cutoff on luminosity to search for anomalies? We test this hypothesis by making such a cut.  We rank each event by peak luminosity (in any filter). If we keep the top 1\% of brightest events, 50\% of the sample is made up of Type Ia SNe and only $\approx 4$\% are SLSNe. We can further make a cut on photometric redshift uncertainties. Even with extremely aggressive cuts ($\sigma_z/z<0.05$) which remove 99.99\% of the sample, only $\approx 10$\% of the sample are SLSNe. The plurality, $\approx 40$\%, are Type Ia SNe. Our detection algorithm, although biased towards brighter events, is sensitive to all outliers regardless of luminosity. 

\subsection{Anomaly Detection in Live Streaming Data}

\begin{figure*}
  \centering
  \includegraphics[width=\textwidth]{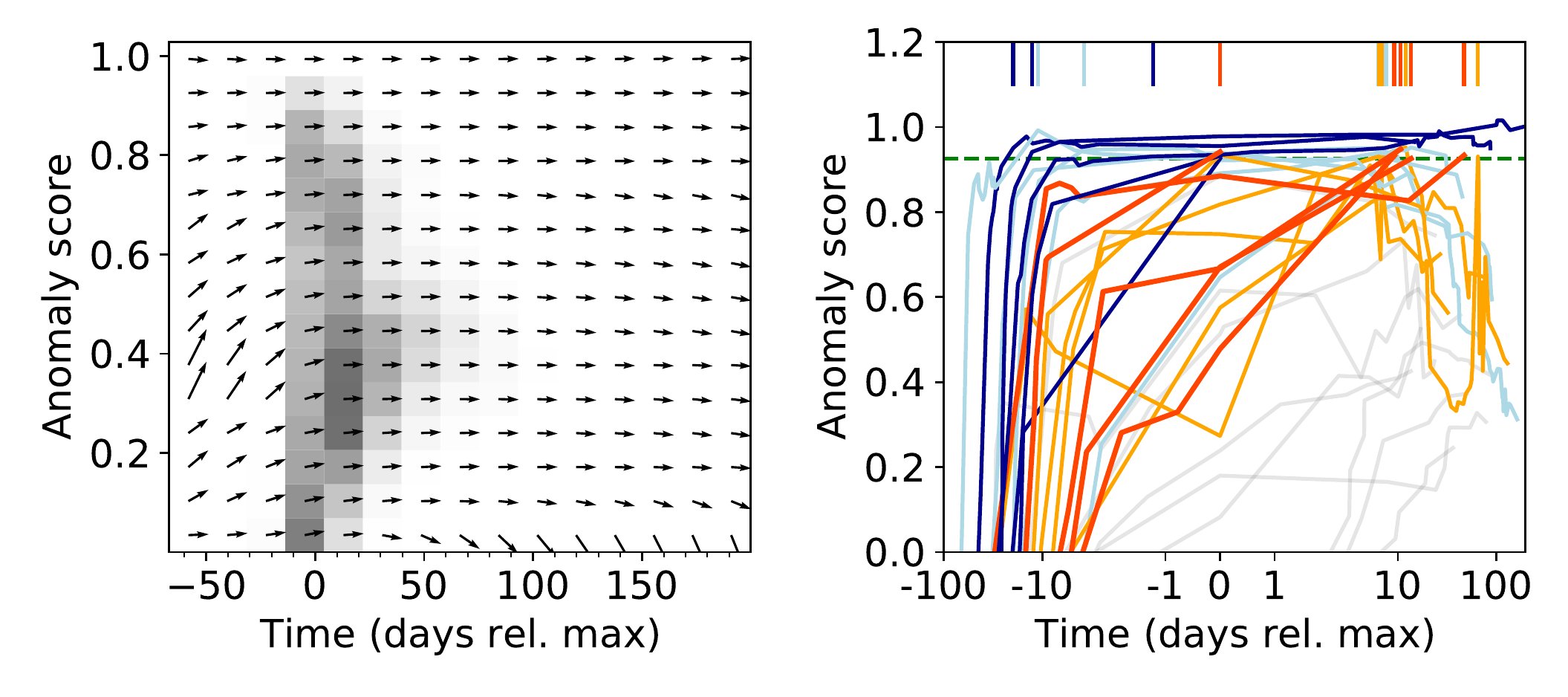}
  \caption{Evolution of anomaly scores as a function of time for a representative set of SNe. \textit{Left:} Vector plot showing the flow of anomaly scores over time. Arrows represent the average gradients of anomaly scores. \textit{Right:} Anomaly score curves for a representative set of SNe. The green dotted line is the 99\textsuperscript{th} percentile threshold for final values (having an anomaly score of $\mathrm{A}\approx0.95$). Grey curves are SNe which never cross the anomaly threshold. Blue and orange curves represent events from the minority and majority classes, respectively. The light orange/blue curves drop below the anomaly threshold before the final observation, while darker curves remain above the anomaly threshold until the end. The colored vertical dashes above the anomaly curves represents the times in which the anomalous events \textit{first} cross the anomaly threshold. Members of the minority class are much more likely to cross this trigger threshold \textit{before} maximum light, while erroneous anomalies from the majority classes are triggered after peak luminosity.  \label{fig:anomtime}}
\end{figure*}

Our analysis thus far has focused on the full light curves, rather than real time followup. We next turn to how the anomaly scores evolve over time for a representative subsample of $\approx10^3$ SNe. As shown in Figure~\ref{fig:anomerr}, most events are identified near peak luminosity ($t\approx 0$). We focus on how anomaly scores vary for the majority classes versus the minority classes. Events from the minority classes are much more likely to be triggered as anomalous before peak. In our representative sample, $\approx 60$\% of Type Ia/Ibc/II SNe identified as anomalous are first marked as such \textit{after} peak magnitude. Type Ia and Type Ibc SNe are typically flagged just around peak, while Type II SNe are flagged, on average, about a week post-peak. In contrast, $\approx 65$\% of anomalies from minority classes are identified as anomalous before peak. They are flagged, on average, about one week before peak. However, short-lived anomalous transients (such as ILOTs and CaRTs) are flagged around or post-peak. We visualize our findings for a representative sample of transients in Figure~\ref{fig:anomtime}.  The background histogram of the vector plot in Figure~\ref{fig:anomtime} shows the overall density of the anomaly scores over time. Most curves have just a few points around $t=0$ and have low anomaly scores. The arrows show that the anomaly scores typically rise before peak but plateau after peak. This implies that, on average, the scores are steady post-peak. 

Finally, we investigate if light curves often begin as anomalous and then drop to less anomalous over time. For this test, we use a 99th-percentile cutoff in anomaly score. We show a selection of representative anomaly curves over time in Figure~\ref{fig:anomtime}; in this figure, grey curves are SNe which never reach the anomaly threshold; blue curves are members of the minority classes which reach the anomaly threshold; and orange curves are members of the majority classes which reach the anomaly threshold. Light blue/orange curves represent members of the minority/majority classes which reach the anomaly threshold but drop below threshold before the end of the event. In contrast, dark curves are those which remain anomalous until the end of the event. We find that about 7\% of events reach this threshold at least once (with, by definition, 1\% remaining anomalous by the final observation). As previously mentioned, we find that members of the minority classes are typically flagged as anomalous before peak, while members of the majority classes are flagged after peak; this is regardless of whether or not those events drop below the anomaly threshold by the final observation. The fact that the minority classes are flagged before peak is useful in practice, as we are much more likely to follow events caught \textit{before} peak luminosity.

\section{Conclusions}

We presented an anomaly detection pipeline for SN-like transients in an LSST-like data stream. We re-purpose the \plasticc\ dataset to train and test our algorithm, allowing us to analyze the \textit{how}, \textit{why},  and \textit{when} events are tagged as anomalous. Our key results are as follows:

\begin{itemize}
    \item We present a novel variational recurrent autoencoder neural network architecture that encodes SN-like light curves in real time into a low-dimension vector.  We train this neural network on 1\% of \Ntrans\ events from the \plasticc\ dataset.
    \item We pair this neural network with an isolation forest to assign every transient an anomaly score. We use a Monte Carlo method to estimate our uncertainty on this score.
    \item We examine the efficacy of our algorithm through a series of percentile and uncertainty cuts. We find that our algorithm is successful in identifying anomalous classes, especially luminous events.
    \item We find that our algorithm is often limited by the photometric redshift estimate. Catastrophically incorrect redshift estimates of Type Ia SNe are especially challenging to remove from our anomaly samples.
    \item We find that members of minority classes (i.e., SNe which are not Type Ia, Type Ibc or Type II) are likely to be identified before peak luminosity. In contrast, erroneously flagged members of the majority classes are more likely to be flagged post-peak.
\end{itemize}

Much is left to be done to sufficiently prepare for the deluge of data which will come in the new era of the Rubin Observatory. This algorithm must be tested on real data with an active follow up campaign. Furthermore, it is possible to use the anomaly score designed here in classification methods to increase the purity of rare transient samples at the cost of completeness.

\acknowledgements
We thank Lehman Garrison for insightful conversations. VAV is supported by the Simons Foundation through a Simons Junior Fellowship (\#718240). The Berger Time Domain group at Harvard is supported in part by NSF and NASA grants, as well as by the NSF under Cooperative Agreement PHY-2019786 (The NSF AI Institute for Artificial Intelligence and Fundamental Interactions \url{http://iaifi.org/}). This research made use of the following software packages:
\texttt{numpy} \citep{numpy},
\texttt{scipy} \citep{scipy},
\texttt{jupyter} \citep{jupyter},
\texttt{sklearn} \citep{sklearn},
\texttt{matplotlib} \citep{matplotlib},
\texttt{tensorflow} \citep{tensorflow}, and
\texttt{astropy} \citep{astropy}.

\bibliography{sample63}{}
\bibliographystyle{aasjournal}

\appendix

\begin{figure*}
  \centering
  \includegraphics[width=\textwidth]{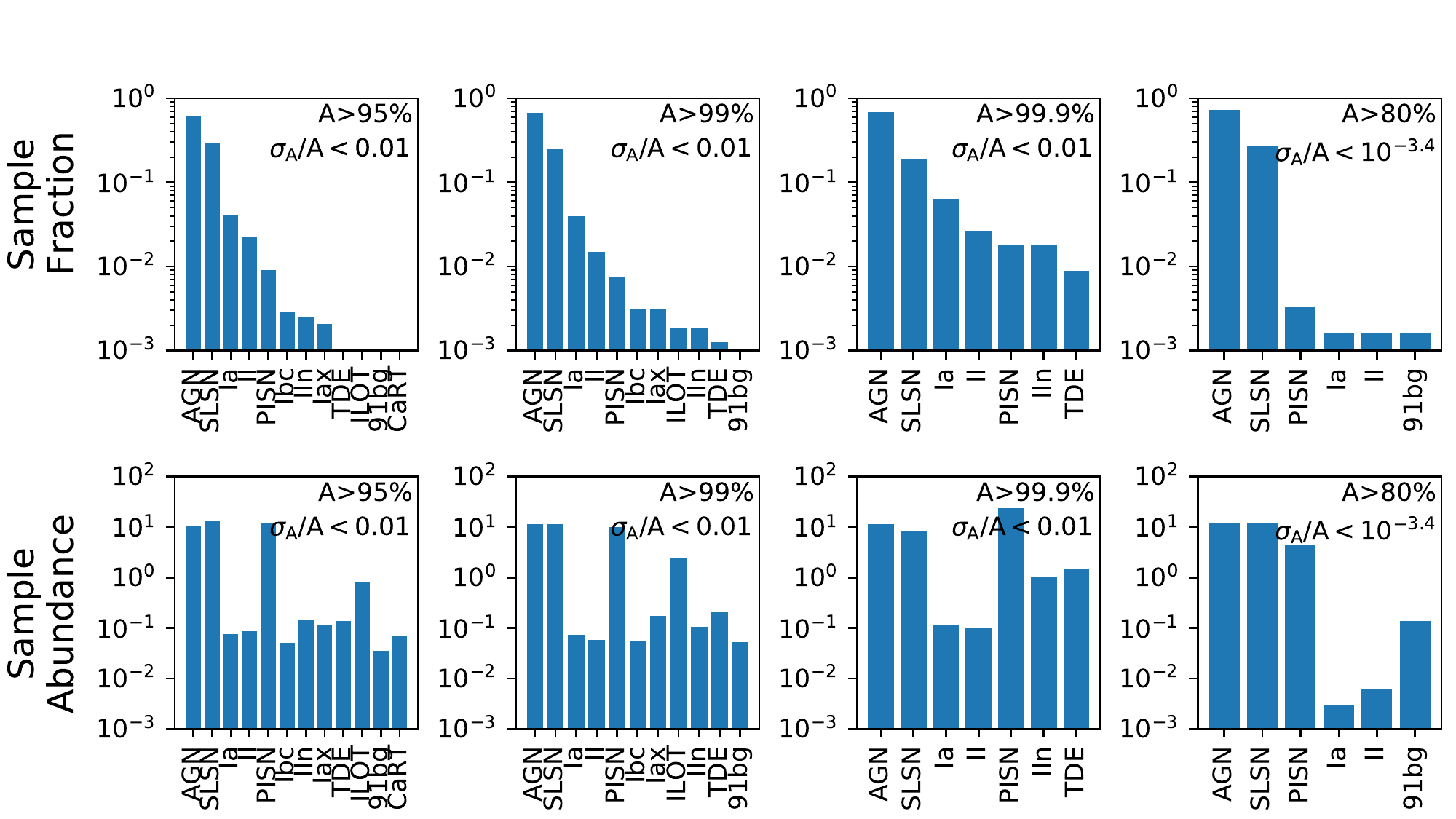}
  \caption{\textit{Top row:} Breakdown of anomalous events, given the cuts listed in each panels. Cuts are identical to those of Figure~\ref{fig:histdist}, but now include AGN. AGN dominate the samples in all cuts. \textit{Bottom row:} Relative abundance of anomaly samples shown in top rows. AGN are, like members of the minority class, highly over-represented in the anomaly sample. \label{fig:histdistagn}}
\end{figure*}

Here we investigate the AGN included in the \plasticc\ simulation.  In Figure~\ref{fig:histdistagn}, we show the same anomaly samples shown in Figure~\ref{fig:histdist} but including AGN. AGN dominate the anomaly sample in all cuts. Our algorithm over-represents AGN in our anomaly sample at the same rate as SLSNe and PISNe.

However, this is not likely not representative of reality. The \plasticc\ AGN are generated via a damped random walk with a structure function to define the correlation between each filter, as described in \cite{macleod2011quasar}. We find that, amongst the AGN identified as ``anomalous'' in this study, all AGN have at least 20 data points spanning at least 60 days in duration. These events would very likely be identified as AGN via other classification methods, and be removed from our anomaly datastream. These events are an example of a simple anomaly detection, in which simple filters would likely also pick them out as distinct from SNe (see a summary of similar pitfalls in \citealt{2020arXiv200913807W}).

Finally, we note that being able to identify AGN in the LSST datastream is an open problem \citep{agnwhitepaper}. At first glance, our algorithm is seemingly very successful at identifying AGN; however, these are likely AGN which will be quickly identified by other means. Future studies will need to identify extreme outbursts of AGN (not explicitly simulated by \plasticc) and AGN with substantial dust extinction (likely not identified in the LSST alert stream unless they undergo a significant outburst). Searching for these elusive AGN continue to be an open problem which require further development of specialized classification methods.

\end{document}